\documentstyle[preprint,prc,aps]{revtex}

\begin{document}
\draft
\title{\bf
The anapole moment and nucleon weak interactions}
\author{V. V. Flambaum and D. W. Murray}
\address{School of Physics, University of New South Wales,
Sydney, 2052, Australia}
\maketitle
\begin{abstract}
{}From the recent measurement of parity nonconservation (PNC) in the
Cs atom we have extracted the constant of the nuclear spin dependent
electron-nucleon PNC interaction, $\kappa = 0.442 (63)$; the anapole
moment constant, $\kappa_a = 0.364 (62)$; the strength of the PNC
proton-nucleus potential, $g_p = 7.3 \pm 1.2 \mbox{ (exp.)} \pm
1.5 \mbox{ (theor.)}$; the $\pi$-meson-nucleon interaction
constant,
$f_\pi \equiv h_\pi^{1} = [9.5 \pm 2.1 \mbox{ (exp.)}
\pm 3.5 \mbox{ (theor.)}] \times 10^{-7}$;
and the strength of the
neutron-nucleus potential, $g_n = -1.7 \pm 0.8 \mbox{ (exp.)}
\pm 1.3 \mbox{ (theor.)}$.
\end{abstract}
\vspace{5mm}
\vspace{10mm}

In the work \cite{science97} the parity
nonconserving (PNC) transition amplitude
between the $6s$ and $7s$ states of the $^{133}$Cs atom
has been precisely
measured:
\begin{equation}
E \equiv - \mbox{Im} (E1_{\mbox{PNC}}) / \beta = 1.5935 (56)
\mbox{mV} / \mbox{cm}.
\label{eo1}
\end{equation}
They also observed the nuclear spin-dependent contribution
\begin{equation}
\mbox{Im} (E1_a) / \beta = 0.077 (11) \mbox{mV} / \mbox{cm}.
\label{eo2}
\end{equation}
This is a manifestation of parity violation in atomic nuclei and
provides
the first measurement of a nuclear anapole moment --- an
electromagnetic
multipole violating the fundamental symmetries of parity and charge
conjugation invariance. The anapole moment was introduced by
Zel'dovich \cite{zeldovich1957}
just after the discovery of parity violation. He pointed
out that a particle should have a parity-violating electromagnetic
form factor, in addition to the usual electric and magnetic
form factors.
The first realistic example, the nuclear anapole moment, was
considered in Ref. \cite{FK1980} and calculated
in Ref. \cite{FKS1984}.
In these works it was also demonstrated that atomic and molecular
experiments could detect anapole moments.
Subsequently, a number of experiments
were performed in Paris, Boulder,
Oxford, and Seattle
\cite{exper} and some limits on the magnitude of the anapole
moment were established. However,
the first unambiguous detection of the
nuclear anapole moment (14\% accuracy) has just been completed
\cite{science97}.

The existence of the anapole moment is due to parity nonconserving
nuclear forces which create spin and
magnetic moment helical structures
inside the nucleus. (A detailed
discussion of the spin helix produced
by the weak interaction is contained in Ref. \cite{Kbook}).
The wave function of the unpaired nucleon can be
presented as
\begin{equation}
\psi = e^{i \theta \mbox{\boldmath $\sigma$}
\cdot {\bf r}} \psi_0,
\label{eex22a}
\end{equation}
i.e., the spin ${\bf s} = \frac{1}{2} \bbox{\sigma}$ is
rotated around
the vector ${\bf r}$. Here the angle of rotation, $\theta r$ is
proportional to the strength of the weak interaction
[$\theta = -\frac{G}{\sqrt{2}} g \rho$,
see Eq. (\ref{eex18})] and $\psi_0$
is the unperturbed wave function. The correction to
the electromagnetic
currents due to this spin rotation has a toroidal structure.
The toroidal electromagnetic
current density ${\bf j}$ produces a magnetic field inside
the torus like that inside a classical toroidal coil. In the
limit of a
point-like nucleus the vector potential corresponding to
this magnetic field can be presented as \cite{FK1980,FKS1984}
\begin{eqnarray}
{\bf A} & = & {\bf a} \delta (r) \nonumber \\
{\bf a} & = & - \pi \int {\bf j} (r) r^2 \, d^3 r
= \frac{1}{e} \frac{G}{\sqrt{2}} \frac{K {\bf I}}{I (I+1)} \kappa_a,
\label{eop2a}
\end{eqnarray}
where ${\bf a}$ is an anapole moment vector
directed along the nuclear
spin ${\bf I}$, $K = (I + \frac{1}{2}) (-1)^{I + 1/2 -l}$
($l$ is the orbital angular momentum of the external nucleon), and
$e$ is the electric charge of the proton. We separated the
Fermi constant of the weak
interaction ($G$) and introduced the
dimensionless constant $\kappa_a$.
The operator of the anapole moment,
$\hat{\bf a}$
(${\bf a} = \langle \psi | \hat{\bf a} | \psi \rangle$)
is given by the following formula \cite{FMDNP87}:
\begin{equation}
\hat{\bf a} = \frac{\pi e}{m} \left[\mu ({\bf r}
\times \bbox{\sigma}) - \frac{q}{2} ({\bf p} {r}^2 + {r}^2 {\bf p})
\right],
\label{eopaa}
\end{equation}
where $m$ is the mass of a nucleon,
${\bf r}$ and ${\bf p}$ are the position and momentum operators
of the nucleon,
$\mu$ is the nucleon magnetic moment in nuclear magnetons
and $q = 0$ ($1$) for a neutron (proton).
The dominant contribution to the nuclear anapole is given by
the spin current [the first term in Eq. (\ref{eopaa})].
The contribution
of the second term (the convection or orbital current
contribution) is very small.
Moreover, to a large extent it is canceled out by the
contribution of the contact current (see Refs.
\cite{FK1980,FKS1984,DKTnp1994}).
The only other sizeable contribution is due to the
spin-orbit current considered in Ref. \cite{DKTnp1994} and is about
$-20 \%$ of the dominant spin contribution.

The interaction between atomic electrons and the magnetic field of
the nuclear anapole produces a nuclear spin dependent PNC effect in
atoms, which was first calculated in Ref. \cite{NSFK1977} and
has been measured in Ref. \cite{science97}.
The PNC amplitudes for different hyperfine
transitions were found to be different.
This difference is produced by the magnetic interaction
of the atomic
electron and the anapole vector-potential ${\bf A}$:
\begin{equation}
V_a = e \bbox{\alpha} \cdot {\bf A} = e \bbox{\alpha} \cdot {\bf a}
\delta (r)
= \frac{G}{\sqrt{2}}  \frac{K {\bf I} \cdot
\bbox{\alpha}}{I (I+1)} \kappa_a \delta (r).
\label{eop2c}
\end{equation}

Note that there are other mechanisms
that produce (small) atomic effects
similar to the anapole moment. This means that the atomic electron's
interaction with the nucleus
should actually be described by Eq. (\ref{eop2c}) with $\kappa_a$
replaced by a new constant, $\kappa$ (more on this below).

Accurate atomic calculations of the PNC effect produced by
the interaction (\ref{eop2c})
have been done in Refs. \cite{frant88,kraft88,BPzphys1991,blund92}.
The result of the many-body calculation in Ref. \cite{kraft88}
is very close to
the semiempirical calculation in Ref. \cite{frant88}.
The result of the Hartree-Fock calculation \cite{blund92} differs
by about $10 \%$ since it does not include many-body corrections.
To reduce the theoretical error we
calculate the ratio of the nuclear spin-dependent PNC amplitude to
the main spin independent PNC amplitude. Using the most complete
many-body calculation of the nuclear spin dependent PNC
amplitude \cite{kraft88}, the calculation of the main PNC amplitude
\cite{DFSS1987} (which was done
using the same method and computer codes) and
the experimental data for different hyperfine transitions from
Ref. \cite{science97} we obtain the following equations:
\begin{eqnarray}
E (1 + 0.05814 \kappa) = & 1.6349(80) \mbox{ mV} / \mbox{cm}
& \;\;\;\mbox{ (for the $6S_{F=4} \rightarrow 7S_{F=3}$ transition)}
\nonumber \\
E (1 - 0.05148 \kappa) = & 1.5576(77) \mbox{ mV} / \mbox{cm} &
\;\;\;\mbox{ (for the $6S_{F=3} \rightarrow 7S_{F=4}$ transition)}
\label{eex21aa}
\end{eqnarray}
The solution to these equations is
\begin{equation}
E = 1.5939 (56) \mbox{mV} / \mbox{cm}
\label{eex21ab}
\end{equation}
\begin{equation}
\kappa = 0.442 (63).
\label{eex21ac}
\end{equation} 
The calculated ratio of the nuclear spin dependent PNC amplitudes
in (\ref{eex21aa}) to the
main PNC amplitude ($E$) is known very accurately, i.e.,
there is practically no theoretical error in the extracted value of
$\kappa$. This value of $\kappa$ contains three contributions:
\begin{equation}
\kappa = \kappa_a - \frac{K-1/2}{K} \kappa_2 + \frac{I+1}{K}
\kappa_Q,
\label{eop3b}
\end{equation}
where $K = 4$ and $I = \frac{7}{2}$ for $^{133}$Cs, 
$\kappa_a$ is the anapole moment contribution (\ref{eop2a}),
$\kappa_2 = 1.25 (2 \sin^2 \theta_W - \frac{1}{2})
\approx -0.05$ is the
contribution of that part of the weak electron-nucleus interaction
that depends on the
nuclear spin (see, e.g., \cite{Kbook,NSFK1977}), and
$\kappa_Q$ is the
contribution of the combined action of the nuclear spin independent
electron-nucleus weak interaction and the hyperfine
interaction \cite{FK1985} (see
also Refs. \cite{kozlov1988,BPpl1991}):
\begin{equation}
\kappa_Q = -\frac{1}{3} Q_W \frac{\alpha \mu_N}{m R_N} =
2.5 \times 10^{-4}
A^{2/3} \mu_N = 0.017.
\label{eop4a}
\end{equation}
Here $Q_W$ is the weak charge of the nucleus,
$\alpha = e^2 = 1/137$,
$R_N = r_0 A^{1/3}$ is the nuclear radius, and $\mu_N$ is the
magnetic moment of the nucleus in nuclear magnetons (for $^{133}$Cs
$\mu_N = 2.58$).
The value of $\kappa_Q$ obtained in the more complete calculation
in Ref. \cite{BPpl1991} is about $1.5$ times larger
(as it contains some average radius of the nucleon distribution,
$\bar{R}$ instead of $R_N$, as in the above equation),
i.e., $\kappa_Q \approx 0.025$.
{}From the above results it
follows that $\kappa_a = 0.370 (63)$.

The Hamiltonian of the electron-nucleon
interaction (\ref{eop2c}) is presented for a
point-like nucleus. However, a real nucleus has a finite size.
Therefore, the ``anapole moment'' measured in the experiment
\cite{science97} is in fact different than the anapole
moment defined in (\ref{eop2a}). The ``anapole moment'' that
was measured in the
experiment can be defined as \cite{FH1993}
\begin{equation}
\tilde{{\bf a}} = - \pi \int {\bf j} (r) r^2
(1 - Z^2 \alpha^2 u(r)) \, d^3 r
\approx (1 - 0.3 Z^2 \alpha^2) {\bf a},
\label{en24aaa}
\end{equation}
where $u(r) \approx \frac{1}{4} \left(\frac{r}{R_N} \right)^2
- \frac{1}{30} \left(\frac{r}{R_N} \right)^4$.
For Cs $Z^2 \alpha^2 = 0.16$. The interaction due to this
``anapole moment'' is just Eq. (\ref{eop2c}) with ${\bf a}$ replaced
by $\tilde{{\bf a}}$, i.e., the ``anapole moment'' is placed
at the center of the nucleus. However, in the previous
atomic calculations \cite{frant88,kraft88,BPzphys1991,blund92} a
different
``regularization'' prescription was used:
$\delta(r)$ was replaced by a finite range function,
$\tilde{\delta}(r)$ that has the shape of the nuclear density.
The electron part of
the anapole moment interaction (\ref{eop2c}) mainly
mixes $s_{1/2}$ and $p_{1/2}$ electron orbitals. Using the electron
wave functions inside the nucleus presented in
Ref. \cite{Kbook} we have
\begin{equation}
\langle \psi_s | e \bbox{\alpha} \cdot \tilde{{\bf a}}
\delta (r) | \psi_p \rangle
= \frac{\langle \psi_s | e \bbox{\alpha} \cdot \tilde{{\bf a}}
\tilde{\delta} (r) | \psi_p \rangle}{1-0.4 Z^2 \alpha^2}
= \frac{1-0.3 Z^2 \alpha^2}{1-0.4 Z^2 \alpha^2}
\langle \psi_s | e \bbox{\alpha} \cdot {\bf a}
\tilde{\delta} (r) | \psi_p \rangle.
\label{ee24aac}
\end{equation}
This means that to accurately
take into account the finite nuclear size
the results of the atomic calculations of the
anapole moment contribution
\cite{frant88,kraft88,BPzphys1991,blund92} should be
multiplied by $(1-0.3 Z^2 \alpha^2)/(1-0.4 Z^2 \alpha^2)
\approx 1 + 0.1 Z^2 \alpha^2 = 1.016$. Therefore
the true value of $\kappa_a$ will be $1.6 \%$ smaller
than $0.37$:
\begin{equation}
\kappa_a = 0.364 (62).
\label{eee24aaad}
\end{equation}
The value $0.36$ has also been obtained in \cite{KhripPC}.

In Ref. \cite{FKS1984} analytical and numerical
calculations of $\kappa_a$ have been done.
The approximate analytical formula was obtained
by using the wave function
(\ref{eex22a}) to calculate the mean value of the anapole moment
operator (\ref{eopaa}). The result is
\begin{equation}
\kappa_a = \frac{9}{10} \frac{\alpha \mu} {m r_0} A^{2/3} g_p
= 0.08 g_p,
\label{eop4c}
\end{equation}
where $\mu$ is the magnetic moment of
the external nucleon in nuclear
magnetons and $r_0 = 1.2 \mbox{ fm}$.
The more accurate numerical calculations \cite{FKS1984,DKTnp1994}
in a Saxon-Woods potential
with a spin-orbit
correction give the following for $^{133}$Cs:
\begin{equation}
\kappa_a = 0.06 g_p.
\label{eop4d}
\end{equation}
Here $g$ is the dimensionless strength constant in the
weak nucleon-nucleus potential:
\begin{equation}
\hat{W} = \frac{G}{\sqrt{2}} \frac{g}{2m} [\bbox{\sigma}
\cdot {\bf p}
\rho(r) + \rho(r) \bbox{\sigma} \cdot {\bf p}],
\label{eex18}
\end{equation}
where $\rho(r)$ is the number density of core nucleons
($g = g_p$ for a proton).

The proton-nucleus and neutron-nucleus constants
can be expressed in terms
of the meson-nucleon parity nonconserving interaction
constants \cite{FKS1984,FPhysScr}
(we use the notation of Ref. \cite{DDH1980}):
\begin{eqnarray}
g_p & = & 2.0 \times 10^5 W_{\rho} \left[ 176
\frac{W_{\pi}}{W_{\rho}} f_{\pi}
-19.5 h_{\rho}^0 - 4.7 h_{\rho}^1 + 1.3 h_{\rho}^2
-11.3 (h_{\omega}^0 + h_{\omega}^1) - 1.7
{h_{\rho}^{1}}^{'} \right] 
\nonumber \\
g_n & = & 2.0 \times 10^5 W_{\rho}
\left[ -118 \frac{W_{\pi}}{W_{\rho}} f_{\pi}
-18.9 h_{\rho}^0 + 8.4 h_{\rho}^1 - 1.3 h_{\rho}^2
-12.8 (h_{\omega}^0 - h_{\omega}^1) + 1.1 {h_{\rho}^{1}}^{'} \right]
\label{eop5a}
\end{eqnarray}
The parameters $W_{\rho}$
and $W_{\pi}$ are present in the above equation
to take into account the nucleon-nucleon repulsion
at small distances and
the finite range of the true interaction potential.
As in Ref. \cite{FKS1984}, we use the calculations
of PNC for neutron and proton
scattering on $^4$He \cite{nphe4}, and take
$W_{\rho} = 0.4$ and $W_{\pi} = 0.16$.
Using the ``best'' values of the $f$ and $h$ constants listed in
Ref. \cite{DDH1980}
(from here on we will refer to these as the DDH ``best'' values)
one obtains $g_p = 4.5$, $g_n = 0.2$,
and $\kappa_a = 0.27$.
Note that this is a single particle shell-model value of the anapole
moment constant. Shell-model calculations usually have an
accuracy of about 30\%.
Thus, the agreement between the experimental value
($0.364 \pm 0.062$) and the theoretical value ($0.27$) is as good
as could be expected.
(Moreover, it was shown in Ref. \cite{FV1994} that
the RPA corrections to the
weak potential increase $g_p$ by 30\%, 
thus $\kappa_a$ could be increased to very close
to the central experimental number of $0.364$.)

Comparing the measured value of $\kappa_a$ (\ref{eee24aaad})
with the theoretical expression (\ref{eop4d}) gives
\begin{equation}
g_p = 6 \pm 1 \mbox{ (exp.)}.
\label{eop6a}
\end{equation}
We do not present here the theoretical error
from the nuclear calculation
of $\kappa_a$ (about $30\%$).

Now we can use the expression for $g_p$ in
terms of the meson-nucleon interaction constants
to find $f_{\pi}$.
It was stated in the recent review \cite{brown}
that experiments give values of the
$\rho$ and $\omega$ weak constants very close to
the DDH ``best'' values (these constants can be
found from, e.g.,  p-p and p-$\alpha$ PNC experiments).
The contribution of $\rho$ and $\omega$ to $g_p$ is
$g_p (\rho, \omega) = 2$.
The main controversy is about the value of
$f_{\pi} \equiv h_{\pi}^{1}$. Comparison between (\ref{eop6a}) and
(\ref{eop5a}) gives
\begin{equation}
f_{\pi} = (g_p - 2) \times 1.8 \times 10^{-7}
= (7 \pm 2 \mbox{ (exp.)}) \times 10^{-7}.
\label{eop6b}
\end{equation}
We stress once more that the theoretical
error in the nuclear calculation of
$\kappa_a$ is ignored here. Then,
using this value of $f_{\pi}$ and the
DDH ``best'' values of $h_{\rho}$ and $h_{\omega}$
in Eq. (\ref{eop5a}) we obtain
\begin{equation}
g_n = -0.38 \times 10^{7} f_{\pi}
+ 1.9 = -0.9 \pm 0.7 \mbox{ (exp.)}.
\label{eop6c}
\end{equation}

There are other nuclear calculations of $\kappa_a$
\cite{HHM1989,BPzphys1991,BPpl1991,DT1997}.
Ref. \cite{HHM1989} contains a detailed calculation of
the $\pi$-meson contribution to the anapole moment and
Refs. \cite{BPzphys1991,BPpl1991} include some configuration
mixing effects.
The most complete calculation of the anapole moment has been done in
Ref. \cite{DT1997}: they included
all single-particle contributions (spin,
spin-orbit, convection, and contact currents) and
many-body corrections
in the RPA approximation (e.g., the induced PNC interaction and the
above mentioned
RPA renormalization of the weak potential, which were considered in
Refs. \cite{FV1994,FV1995}). For comparison, it
is convenient to present the result
of their calculation in a form that stresses
the role of $g_p$:
\begin{equation}
\kappa_a = 0.05 (g_p + 0.16 g_n - 0.07 g_{pp} - 0.01 g_{np}),
\label{eex20a}
\end{equation}
where $g_{pp}$ and $g_{np}$ are the
constants of the proton-proton and
neutron-proton weak interactions;
these are related to $g_p$ and $g_n$ by the
formulae $g_p = (Z/A) g_{pp} + (N/A) g_{pn}$ and
$g_n = (Z/A) g_{np} + (N/A) g_{nn}$
(see Ref. \cite{ST1993}).
The authors of Ref. \cite{DT1997}
estimated the theoretical error in (\ref{eex20a})
as smaller than $20\%$. For the DDH ``best'' values
of the meson-nucleon
weak interaction constants we have $g_n = 0.2$, $g_{pp} = 1.5$, and
$g_{np} = -2.2$ \cite{ST1993} and so we obtain
\begin{equation}
\kappa_a = 0.05 (g_p - 0.05).
\label{eex20b}
\end{equation}
Thus, to an accuracy of $\sim 1\%$ $\kappa_a$ is still proportional
to $g_p$. Comparing this with the experimental value of $\kappa_a$
in Eq. (\ref{eee24aaad}) we obtain
\begin{equation}
g_p = 7.3 \pm 1.2 (\mbox{exp.}) \pm 1.5 (\mbox{theor.}).
\label{eex20c}
\end{equation}
Once again we can use the value of $g_p$ to find a value of
$f_{\pi}$. Comparing the expression for $g_p$
(\ref{eop5a}) with its numerical value (\ref{eex20c}) we obtain
\begin{equation}
f_{\pi} \equiv  h_\pi^{1} = [9.5 \pm 2.1 \mbox{ (exp.)}
\pm 3.5 \mbox{ (theor.)}] \times 10^{-7}.
\label{eex20d}
\end{equation}
We increased the theoretical error here from $2.7$ to $3.5$ to take
into account the uncertainty in the relation between $g_p$ and
$f_{\pi}$ (\ref{eop5a}).
As before, we use this
value of $f_{\pi}$ and the DDH ``best'' values
of $h_{\rho}$ and $h_{\omega}$ in Eq. (\ref{eop5a}) and we obtain
\begin{equation}
g_n = -1.7 \pm 0.8 \mbox{ (exp.)} \pm 1.3 \mbox{ (theor.)}.
\end{equation}

We have presented two sets of estimates of $g_p$, $f_{\pi}$,
and $g_n$ to give an indication of the possible spread
of the results due to theoretical uncertainty. These
two sets of results agree with each other to within their errors.
In the abstract we presented values based on the more complete
many-body calculations.
 
Now we will compare our estimates of $f_{\pi}$,
Eqs. (\ref{eop6b}) and (\ref{eex20d}),
with other estimates in the literature.
There is no contradiction between these values of
$f_{\pi}$ and the QCD calculations, which give $f_{\pi} \equiv
h_\pi^{1} = 5
\mbox{--} 6 \times 10^{-7}$ \cite{khatsimovskii1985,kaplan1993}.
The DDH ``best'' value of
$f_{\pi}$ is $f_{\pi} = 4.6 \times 10^{-7}$.
However, there are also smaller estimates of
$f_{\pi}$ in the literature,
going down to the value $|f_{\pi}| < 1.3 \times 10^{-7}$
derived from a $^{18}$F PNC measurement
(see, e.g., the review \cite{Desplanques}).

Note that there could also
be a more exotic interpretation of the results
of the $\kappa$ measurement:
$\kappa_2$ may not be described by the standard electroweak theory
and so may have a larger magnitude, thus implying a smaller
value of $\kappa_a$ [see Eq. (\ref{eop3b})], and hence $f_{\pi}$.
However, such an explanation
would be very improbable since the results of
measurements of atomic weak charges and PNC in deep inelastic
electron-nucleon scattering agree with the standard model.
To clear this question
it would be interesting to measure the anapole moment of
the $^{207}$Pb nucleus, which contains
an external neutron. The constant
$g_n$ contains $f_{\pi}$ with a negative sign
in this case [see Eq. (\ref{eop5a})].

Just before the submission of this paper it was brought to
our attention that an
analysis of nucleon weak interactions, based on the experiment
\cite{science97}, has also been
done in the recent work \cite{haxton1997}.

One of us (VVF) is grateful to I.B. Khriplovich for
useful comments.
This work was supported by the Australian Research Council.


\begin{thebibliography}{99}
\bibitem{science97}
C.S. Wood, S.C. Bennett, D. Cho, B.P. Masterson, J.L. Roberts,
C.E. Tanner, and C.E. Wieman, Science {\bf 275}, 1759 (1997).
\bibitem{zeldovich1957}
Ya.B. Zel'dovich,
Zh. \'{E}ksp. Teor. Fiz. {\bf 33}, 1531 (1957)
[Sov. Phys. JETP {\bf 6}, 1184 (1957)].
\bibitem{FK1980}
V.V. Flambaum and I.B. Khriplovich, Zh. \'{E}ksp. Teor. Fiz.
{\bf 79}, 1656 (1980) [Sov. Phys. JETP {\bf 52}, 835 (1980)].
\bibitem{FKS1984}
V.V. Flambaum, I.B. Khriplovich, and O.P. Sushkov, Phys. Lett. B
{\bf 146}, 367 (1984).
\bibitem{exper}
M.A. Bouchiat, J. Gu\'{e}na, L. Pottier, and L. Hunter,
Phys. Lett. B {\bf 134}, 463 (1984);
S.L. Gilbert and C.E. Wieman, Phys. Rev. A {\bf 34}, 792 (1986);
M.C. Noecker, B.P. Masterson and C.E. Wieman, Phys. Rev. Lett.
{\bf 61}, 310 (1988);
N.H. Edwards, S.J. Phipp, P.E.G. Baird, and S. Nakayama,
Phys. Rev. Lett. {\bf 74}, 2654 (1995);
P.A. Vetter, D.M. Meekhof, P.K. Majumder, S.K. Lamoreaux, and
E.N. Fortson, Phys. Rev. Lett. {\bf 74}, 2658 (1995).
\bibitem{Kbook}
I.B. Khriplovich, {\em Parity Nonconservation in
Atomic Phenomena}\/ (Gordon and Breach, Philadelphia, 1991).
\bibitem{FMDNP87}
V.V. Flambaum, in {\em Modern Developments in Nuclear Physics},
edited by O.P. Sushkov (World Scientific,
Singapore, 1987), p. 556.
\bibitem{DKTnp1994}
V.F. Dmitriev, I.B. Khriplovich, and V.B. Telitsin,
Nucl. Phys. A {\bf 577}, 691 (1994).
\bibitem{NSFK1977}
V.N. Novikov, O.P. Sushkov, V.V. Flambaum, and I.B. Khriplovich,
Zh. \'{E}ksp. Teor. Fiz. {\bf 73}, 802 (1977)
[Sov. Phys. JETP {\bf 46}, 420 (1977)].
\bibitem{frant88}
P.A. Frantsuzov and I.B. Khriplovich,
Z. Phys. D {\bf 7}, 297 (1988).
\bibitem{kraft88}
A.Ya. Kraftmakher, Phys. Lett. A {\bf 132}, 167 (1988).
\bibitem{BPzphys1991}
C. Bouchiat and C.A. Piketty, Z. Phys. C {\bf 49}, 91 (1991).
\bibitem{blund92}
S.A. Blundell, J. Sapirstein, and W.R. Johnson,
Phys. Rev. D {\bf 45},
1602 (1992).
\bibitem{DFSS1987}
V.A. Dzuba, V.V. Flambaum, P.G. Silvestrov, and O.P. Sushkov,
J. Phys. B {\bf 20}, 3297 (1987).
\bibitem{FK1985}
V.V. Flambaum and I.B. Khriplovich,
Zh. \'{E}ksp. Teor. Fiz. {\bf 89}, 1505 (1985)
[Sov. Phys. JETP {\bf 62}, 872 (1985)].
\bibitem{kozlov1988}
M.G. Kozlov, Phys. Lett. A {\bf 130}, 426 (1988).
\bibitem{BPpl1991}
C. Bouchiat and C.A. Piketty, Phys. Lett. B {\bf 269}, 195 (1991);
erratum {\bf 274}, 526 (1992).
\bibitem{FH1993}
V.V. Flambaum and C. Hanhart, Phys. Rev. C {\bf 48}, 1329 (1993).
\bibitem{KhripPC}
I.B. Khriplovich (private communication).
\bibitem{FPhysScr}
V.V. Flambaum, Physica Scripta {\bf T46}, 198 (1993).
\bibitem{DDH1980}
B. Desplanques, J.F. Donoghue, and B.R. Holstein, Annals of Physics
{\bf 124}, 449 (1980).
\bibitem{nphe4}
V.F. Dmitriev, V.V. Flambaum, O.P. Sushkov, and V.B. Telitsin,
Phys. Lett. B {\bf 125}, 1 (1983);
V.V. Flambaum, V.B. Telitsin, and O.P. Sushkov,
Nucl. Phys. A {\bf 444},
611 (1985).
\bibitem{FV1994}
V.V. Flambaum and O.K. Vorov, Phys. Rev. C {\bf 49}, 1827 (1994).
\bibitem{brown}
B. Alex Brown, in {\em Parity and Time Reversal Violation in
Compound Nuclear States and Related Topics},
edited by N. Auerbach and
J.D. Bowman (World Scientific, Singapore, 1996), p. 198.
\bibitem{HHM1989}
W.C. Haxton, E.M. Henley, and M.J. Musolf,
Phys. Rev. Lett. {\bf 63},
949 (1989).
\bibitem{DT1997}
V.F. Dmitriev and V.B. Telitsin,
Nucl. Phys. A {\bf 613}, 237 (1997).
\bibitem{FV1995}
V.V. Flambaum and O.K. Vorov, Phys. Rev. C {\bf 51}, 1521 (1995).
\bibitem{ST1993}
O.P. Sushkov and V.B. Telitsin, Phys. Rev. C {\bf 48}, 1069 (1993).
\bibitem{khatsimovskii1985}
V.M. Khatsimovskii, Yad. Fiz. {\bf 42}, 1236 (1985)
[Sov. J. Nucl. Phys. {\bf 42}, 781 (1985)].
\bibitem{kaplan1993}
D.B. Kaplan and M.J. Savage, Nucl. Phys. A {\bf 556}, 653 (1993).
\bibitem{Desplanques}
B. Desplanques, in {\em Parity and Time Reversal Violation in
Compound Nuclear States and Related Topics},
edited by N. Auerbach and
J.D. Bowman (World Scientific, Singapore, 1996), p. 98.
\bibitem{haxton1997}
W.C. Haxton, Science {\bf 275}, 1753 (1997).
\end{thebibliography}
\end{document}